\documentclass[sn-mathphys,Numbered]{sn-jnl}

\usepackage{graphicx}%
\usepackage{multirow}%
\usepackage{amsmath,amssymb,amsfonts}%
\usepackage{amsthm}%
\usepackage{mathrsfs}%
\usepackage[title]{appendix}%
\usepackage{xcolor}%
\usepackage{textcomp}%
\usepackage{manyfoot}%
\usepackage{booktabs}%
\usepackage{algorithm}%
\usepackage{algorithmicx}%
\usepackage{algpseudocode}%
\usepackage{listings}%

\newcommand{\hide}[1]{}

\begin{document}

\title[Ensuring connectedness for the Maximum
Quasi-clique and Densest $k$-subgraph problems]{Ensuring connectedness for the Maximum
Quasi-clique and Densest $k$-subgraph problems}

\author*[1]{\fnm{Daniela} \sur{S. dos Santos}}\email{dssantos@dei.uc.pt}

\author[2]{\fnm{Kathrin} \sur{ Klamroth}}\email{klamroth@uni-wuppertal.de}

\author[3]{\fnm{Pedro} \sur{Martins}}\email{pmartins@iscac.pt}

\author[1]{\fnm{Luís} \sur{Paquete}}\email{paquete@dei.uc.pt}

\affil*[1]{\orgdiv{CISUC, Department of Informatics Engineering}, \orgname{University of Coimbra}, \orgaddress{\street{Pólo II},  \postcode{3030-290}, \city{Coimbra}, \country{Portugal}}}

\affil[2]{\orgdiv{School of Mathematics and Natural Sciences}, \orgname{University of Wuppertal}, \orgaddress{\postcode{42119}, \city{Wuppertal}, \country{Germany}}}

\affil[3]{\orgdiv{Coimbra Business School - ISCAC}, \orgname{Polytechnic Institute of Coimbra}, \orgaddress{\ \postcode{3045-601}, \city{Coimbra}, \country{Portugal}}}

\abstract{Given an undirected graph $G$, a quasi-clique is a subgraph of $G$ whose density is at least $\gamma$ $(0 < \gamma \leq 1)$. Two optimization problems can be defined for quasi-cliques: the Maximum Quasi-Clique (MQC) Problem, which finds a quasi-clique with maximum vertex cardinality, and the Densest $k$-Subgraph (DKS) Problem, which finds the densest subgraph given a fixed cardinality constraint.
Most existing approaches 
to solve both problems
often disregard the requirement of connectedness, which may lead to solutions containing isolated components that are meaningless for many real-life applications.
To address this issue, we propose two flow-based connectedness constraints to be integrated into known Mixed-Integer Linear Programming (MILP) formulations for either MQC or DKS problems. 
We compare the performance of MILP formulations enhanced with our connectedness constraints 
in terms of both
running time and 
number of solved instances against existing 
approaches that
ensure quasi-clique connectedness. Experimental results demonstrate that our constraints are quite competitive, 
making them valuable 
for practical applications requiring connectedness.}



\maketitle

\section{Introduction}\label{sec1}

\label{Int}

Given a simple undirected graph $G = (V, E)$ where $V$ is the set of vertices and
$E$ is the set of edges, a subset of vertices $S$ is called a \emph{clique} if
every two vertices in $S$ are adjacent. The \emph{Maximum Clique Problem} is a
classical combinatorial optimization problem which asks for a clique with the
maximum vertex cardinality in a graph $G$. 

The requirement that every two vertices in a clique are directly linked 
is too restrictive
for many real-life applications that consider graphs from massive,
incomplete, and error-prone data sets, resulting in erroneously missing or
added edges~\cite{Balasundaram2013,Brunato2008}. This has led to the 
definition of different clique relaxations that aim at relaxing particular
properties of cliques such as reachability (diameter), familiarity (degree),
density (edge density), and robustness (connectivity)~\cite{Pattillo2012}. In
this work, we consider a density-based clique relaxation called
\emph{quasi-clique}, which is any subgraph of $G$ whose density is at least a given threshold $\gamma \in (0,1]$.

Two optimization problem variants can be defined for
quasi-cliques~\cite{Abello1999,Balasundaram2013}: $(i)$ given a graph and $\gamma$, 
the problem is to find a $\gamma$-quasi-clique 
with the largest number of vertices; $(ii)$ given a graph and a vertex cardinality $k$,
the problem is to find a subgraph of $k$ vertices with the largest density. 
The variant $(i)$
is called the \emph{Maximum Quasi-Clique} (MQC) Problem while $(ii)$  
is commonly referred to under different terms in the literature, such as \emph{Maximum Edge Subgraph Problem}~\cite{Asahiro2000}, \emph{Densest k-Set Problem}~\cite{Chang2014}, \emph{k-Cluster Problem}~\cite{Corneil1984}, \emph{Heaviest
Unweighted Subgraph Problem}~\cite{Kortsarz1993}; in this work, we refer to it as the \emph{Densest k-Subgraph} (DKS) Problem~\cite{Feige1997}. If DKS requires connectedness, it becomes the Densest Connected $k$-Subgraph (DCKS) Problem~\cite{CHEN2017}.
Since a subgraph of $k$ vertices  
with density $\gamma^*$ for the DKS problem is a feasible 
$\gamma^*$-quasi-clique for the MQC problem with $\gamma = \gamma^*$, 
we will state that any feasible solution for the DKS problem 
is also a quasi-clique.

The MQC and (connected) DKS problems are NP-hard~\cite{Asahiro1995,Feige1997,Patillo2013} and arise in many real-world applications, including social networks~\cite{Brunato2008}, telecommunications~\cite{Abello2002}, and Bioinformatics~\cite{Althaus2014,Backes2011,Bhattacharyya2009}.
Therefore, many exact \cite{Althaus2014,Billionnet2005,Bourgeois2013,Chang2014,Marinelli2021,Pajouh2014,Patillo2013,Ribeiro2019,Veremyev2016}, and heuristic algorithms~\cite{Abello1999,Abello2002,Bhattacharyya2009,Chen2021,Djeddi2019,Macambira2002,Peng2021,Pinto2021,Pinto2018,Zhou2020} have been designed for addressing these problems. Approximation algorithms have also been proposed to tackle the DKS problem~\cite{Bhaskara2010,Bourgeois2013,CHEN2017,Feige2001,hochbaum1994,Kortsarz1993,Liazi2008}.

However, most of the approaches reported in the literature focus on the versions that do not guarantee the connectedness of the resulting subgraphs. 
Not ensuring this property may
lead to subgraphs with isolated components, which are  meaningless 
for applications requiring connectedness, such as those arising in the community
detection~\cite{Fortunato2010} and finding cohesive
clusters~\cite{Gschwind2015}. 
Existing exact
methods ensuring connectedness for the MQC Problem include the Connected$_\mu$ algorithm~\cite{Komusiewicz2015} and Marinelli's connectedness constraints~\cite{Marinelli2021}. Nevertheless, the Connected$_\mu$ algorithm is only capable of effectively addressing the problem for $\gamma \geq 0.5$ whereas Marinelli's approach requires significant computing resources, both in terms of memory and run-time, for large size graphs. 
Concerning the DCKS Problem, Althaus et al.'s lazy constraints~\cite{Althaus2014} ensure connected solutions but may be less efficient when addressing instances that are likely to contain many 
disconnected subgraphs, such as those found in real-life sparse graphs. Finally, the FixCon approach~\cite{Komusiewicz2020} effectively handles the problem only for $k\leq20$.

In order to overcome the limitations of existing approaches, this work
proposes two distinct sets of connectedness constraints: C-STree, which forces the solution to satisfy a spanning tree characterization using a single commodity flow model 
and is applicable to both the MQC and DCKS problems; 
and C-Flow, which is designed specifically for the DCKS problem and relies on a classic flow-based
approach.  
These constraints can be
integrated into well-known MILP models
proposed by Veremyev et al.~\cite{Veremyev2016} and by Billionnet~\cite{Billionnet2005}, 
for MQC and DKS problems, respectively. 

The remainder of this work is organized as follows. 
Section \ref{DefNot} gives the relevant notations and definitions for the scope of this work.
Section \ref{MILP} provides the well-known MILP models for the MQC and DKS problems.
The related works are described in Section \ref{RelWor}. 
Section \ref{ProCons} presents our set of proposed connectedness constraints. 
Section \ref{Res} shows the computational experiments and results. 
Finally, the conclusions are given in Section \ref{Con}.

\section{Definitions and notations}
\label{DefNot}

We consider an undirected and simple graph $G = (V,E)$, where 
$V$ and $E$ are  the vertex and edge set of $G$, respectively. 
For a set of vertices $S \subseteq V$, we denote by $G_S = (V(G_S),E(G_S))$ the subgraph induced by $S$ in $G$ (i.e., $V(G_S)=S$).  Let $A = \{(i,j),(j,i) : \{i,j\} \in E\}$ represent the oriented version of $G$, where each edge $\{i,j\}$ in the original graph is represented by the two bi-directed arcs, $(i, j)$ and $(j, i)$.
The \emph{density} of $G_S$ is denoted by $dens(G_S) = \frac{2\cdot|E(G_S)|}{|V(G_S)|\cdot(|V(G_S)|-1)}$ if $|V(G_S)|\geq 2$. In other words, it is the ratio between $|E(G_S)|$ and the number of edges in a complete graph with $|V(G_S)|$ vertices. If $|V(G_S)| = 1$, we assume that
$dens(G_S) = 1$. 
An induced subgraph $G_S$ is \emph{connected} if any two of its vertices are joined by a path,
otherwise, it is said to be \emph{disconnected}.
A \emph{spanning tree} $T \subseteq G_S$ is a connected spanning subgraph 
with $V(G_S)$ vertices and $|V(G_S)|-1$ edges.

The MQC and DKS problems are formally stated as follows.

\paragraph{Maximum Quasi-clique problem:\\}
\emph{Instance}: A graph $G = (V, E)$ and a constant $\gamma$, where $0 < \gamma \leq 1$.\\
\emph{Problem}: Find an induced subgraph $G_S$ of $G$ such that
$$|V(G_S)| = \max \left\{|V(G_{S^\prime})| : S^\prime \subseteq V \mbox{ and } dens(G_{S^\prime}) \geq \gamma \right\}$$

\paragraph{Densest $k$-subgraph problem:\\}
\emph{Instance}: A graph $G = (V, E)$ and a positive integer $k\leq |V|$. \\
\emph{Problem}: Find an induced subgraph $G_S$ of $G$ such that   
$$dens(G_S)=\max \left\{dens(G_{S^\prime}) : S^\prime \subseteq V \mbox{ and } |V(G_{S^\prime})|=k\right\}$$

When the induced subgraph $G_S$ is required to be connected, MQC becomes the \emph{Maximum Connected Quasi-clique} (MCQC) Problem and DKS becomes the \emph{Densest Connected $k$-subgraph} (DCKS) Problem.
Note that while MQC always has a solution (which may be a single node), the DCKS problem may be infeasible when $G$ is not connected, e.g., when $k=|V|$.

\section{Mixed Integer Linear Programming Models}
\label{MILP}

To the best of our knowledge, the reference MILP models for DKS and MQC problems are those described in~\cite{Billionnet2005} and~\cite{Marinelli2021,Patillo2013,Veremyev2016}, respectively. 
In this work, we evaluate the proposed set of connectedness constraints on 
Model M1~\cite{Billionnet2005} and F3~\cite{Veremyev2016}.
In these formulations, both problems are modelled as an MILP by introducing binary variables for each vertex, 
which indicate whether or not that vertex is included in the maximum quasi-clique. Additionally, the models also include edge binary variables, indicating the edges in the solution. 
The objective function in model M1 is to maximize edge density for the DKS problem, 
while in model F3, the goal is to maximize vertex 
cardinality for the MQC problem. 

In the following, we introduce model M1.

\begin{subequations}\label{eq:M1}
  \begin{align}
            \max \quad &\sum_{\{i,j\}\in E}y_{ij} \nonumber\\
      \text{s.\,t.} \quad 
       & \sum_{i \in V} x_{i}=k \\
       & y_{ij}  \leq  x_{i}  && \forall \{i,j\} \in E  \\
       & y_{ij} \leq x_{j}   &&  \forall \{i,j\} \in E\\
       & y_{ij} \geq  0      && \forall \{i,j\} \in E \\
       & x_{i} \in\{0,1\}    && \forall \ i \in V 
\end{align}
\end{subequations}
where $x_{i}$ and $y_{ij}$ are variables defined for each vertex $i \in V$ and for each edge $\{i,j\} \in E$, respectively,
with $x_{i}=1$ if the vertex $i$ is in the quasi-clique and $y_{ij}=1$ if edge $\{i,j\}$ is in the
quasi-clique. 
The objective function maximizes the number of edges of the quasi-clique.
Constraint (1a) ensures that the cardinality of the quasi-clique is equal to $k$. 
Constraints (1b) and (1c) state that if an edge $\{i,j\}$ is in the quasi-clique, 
then both vertices $i$ and $j$ must be chosen.
Model M1 requires $\mathcal{O}(|V|+|E|)$ variables and $\mathcal{O}(|E|)$ constraints.

In Model F3, the variables $x_{i}$ and $y_{ij}$ are also used. Moreover, a new variable 
$z_{k}$ is introduced to determine the size of the quasi-clique, that is, $z_{k}=1$ if 
its size is $k$. The formulation is given as follows. 

\begin{subequations}\label{eq:F3}
  \begin{align}
            \max \quad &\displaystyle\sum_{i \in V}x_i \nonumber\\
      \text{s.\,t.}  \quad
       &\sum_{\{i,j\}\in E}y_{ij} \geq \gamma \cdot \sum_{k=\ell}^{u}\frac{k \cdot (k-1)}{2} \cdot z_{k}&& \\
       & \sum_{i \in V}x_i = \sum_{k=\ell}^{u} k \cdot z_{k}&&\\
       & \sum_{k=\ell}^{u} z_{k} =1&&\\
       & y_{ij}  \leq  x_{i}&& \forall \ \{i,j\} \in E\\
       & y_{ij}  \leq x_{j}&& \forall \ \{i,j\} \in E\\
       & z_{k} \geq 0       &&  \forall \ k \in \{\ell,...,u\}\\
       & y_{ij} \geq  0      && \forall \ \{i,j\} \in E \\
       & x_{i} \in\{0,1\}    && \forall \ i \in V 
   \end{align}
  \end{subequations}
where $u$ and $\ell$ are upper and lower bounds, respectively, on the size of 
a maximum quasi-clique for a given parameter $\gamma$.
If there is no prior information about the sizes of the quasi-clique, $u$ and $\ell$
can be set to $|V|$ and $1$, respectively.
The objective function maximizes the number of vertices. 
Constraint (2a) ensures the edge density requirements according to the
given parameter $\gamma$. 
Constraints (2b) and (2c) define the cardinality of the quasi-clique.
Note that variable $z_{k}$ does not need to be binary (constraints (2f)). 
This relaxation is supported by Proposition~$1$ in ~\cite{Veremyev2016}, whose proof demonstrates that there exists an optimal solution in which $z_k$ is a binary vector. 
The remaining constraints are the same as for Model M1.

\section{Related Work}
\label{RelWor}

In this Section, we describe the strategies used by the state-of-the-art approaches for MCQC and DCKS problems to guarantee the connectedness of the resulting subgraphs. 

Connected$_\mu$ is a combinatorial branch-and-bound algorithm for the MCQC problem that implements two upper bounding procedures, several data reduction techniques and
termination criteria~\cite{Komusiewicz2015}. This approach leverages the quasi-heredity property of quasi-cliques, 
which states that the iterative removal of the minimum-degree vertex from a quasi-clique will always preserve at least the same density in the remaining subgraphs~\cite{Patillo2013}. This property is used to systematically explore and discover larger connected $\gamma-$quasi-cliques. However, the algorithm can only ensure connected solutions for $\gamma \geq 0.5$ since the quasi-heredity property does not hold for $\gamma < 0.5$.

More recently, Marinelli et al.~\cite{Marinelli2021} proposed a set of 
constraints to ensure connectedness within Model F3.
To this end, the authors introduced the binary variable $c_{i} \in \{0,1\}$ 
for each vertex $i \in V$, where $c_{i}=1$ if and only if vertex $i$ is selected as source vertex, 
and flow variable $f_{ij} \in \mathbb{R}$ for each edge $\{i,j\} \in E$. Therefore, optimal connected
quasi-cliques can be found by adding to Model F3 the following set of constraints.

\begin{subequations}\label{eq:MPR}
  \begin{align}
        &\sum_{i \in V} c_{i} = 1&\\
        & c_{i}  \leq  x_{i} & \forall \ i \in V&\\
        &\sum_{j \in N(i):i<j}f_{ij} - \sum_{j\in N(i):j<i}f_{ji} 
         \geq  \sum_{h \in V} x_{h} - 1 - u(1-c_{i}) & \forall \ i \in V&\\ 
        &\sum_{j \in N(i):i<j}f_{ij} - \sum_{j\in N(i):j<i}f_{ji} 
         \leq  \sum_{h \in V} x_{h} - 1 + u(1-c_{i}) & \forall \ i \in V&\\ 
        &\sum_{j \in N(i):i<j}f_{ij} - \sum_{j\in N(i):j<i}f_{ji} 
         \geq  -1-u(1+c_{i}-x_{i}) & \forall \ i \in V&\\ 
        &\sum_{j \in N(i):i<j}f_{ij} - \sum_{j\in N(i):j<i}f_{ji} 
         \leq  -1 + u(1+c_{i}-x_{i}) & \forall \ i \in V&\\ 
       & f_{ij}  \geq  -(u-1)y_{ij} & \forall \ \{i,j\} \in E&\\
       & f_{ij}  \leq  (u-1)y_{ij} & \forall \ \{i,j\} \in E&\\
       & c_{i} \in\{0,1\}    & \forall \ i \in V&
   \end{align}
  \end{subequations}
where $N(i)$ is the set of neighbors of $i$. 
Constraints (3a) and (3b) ensure that one vertex is selected as the source among those 
belonging to the quasi-clique.
Constraints (3c) and (3d) enforce that 
the number of units of 
flow that leaves the source is the number of vertices in the quasi-clique
minus one, whereas (3e) and (3f) ensure that a single 
unit of flow is absorbed by any other vertices of the quasi-clique except the source node 
$i$ to which $c_i=1$. Constraints (3g) and (3h) set bounds for variables $f_{ij}$, where $f_{ij}$ must be zero if edge $\{i,j\}$ is not in the quasi-clique. 
Finally, Constraint (3i) assures integrality on the $c_i$ variables.
This model requires the addition of $O(|V|+|E|)$ variables and constraints to model F3. We will refer to it as \emph{MPR Model}.

Althaus et al.~\cite{Althaus2014} proposed an approach employing lazy constraints to ensure the connectedness of the solutions for the weighted version of the DCKS Problem. Their proposed constraints enforce that, for each connected component $C \subseteq V$ within the solution where $|C|<k$, at least one neighboring vertex, located outside $C$, is also selected. The mathematical representation of this constraint is as follows:
\begin{subequations}\label{eq:Lazy}
  \begin{align}
        &\sum_{i \in N(C)} x_{i} \geq   x_{j} & \forall \ j \in C, \qquad \forall \ C \subseteq V
   \end{align}
  \end{subequations}
where $N(C)$ denotes the set of vertices in $V$ but not in $C$ with at least one edge incident to a vertex in $C$. 
Constraints from \eqref{eq:Lazy} are then included for subsets $C\subseteq V$ whenever a computed feasible solution is non-connected w.r.t.\ subset $C$.
It is worth noticing that when feasible solutions are more likely to be connected, using lazy constraints might result in faster optimization since the solver focuses on ensuring connectedness only when a disconnected solution is detected. However, if the solution space frequently includes disconnected subgraphs, which is common in sparse graphs, the cost of repeatedly applying lazy constraints might exceed the advantages.

FixCon is a generic solver for the Connected Fixed-Cardinality Optimization Problem (CFCO)~\cite{Komusiewicz2020}. 
In this problem class, the goal is to identify a connected subgraph of size $k$ that maximizes a defined objective function. Specifically, FixCon addresses the DCKS, which is one of the problems falling under the fixed-cardinality category. This approach consists of a subgraph enumeration algorithm along with various generic pruning rules, in addition to a generic heuristic to calculate a lower bound for the objective value. The authors highlight that the theoretical running time guarantees for CFCO are favorable when dealing with small values of $k$. Consequently, FixCon was designed to address the problem for $k\leq20$.

\section{New Connectedness Constraints}
\label{ProCons}
This section introduces two sets of constraints: \emph{C-STree} and \emph{C-Flow}. \emph{C-STree} ensures connected optimal subgraphs for both MQC and DKS 
when integrated into Model F3 and Model M1. This is accomplished by employing a spanning tree characterization based on the single commodity flow 
model proposed by Gavish~\cite{Gavish1982}. 
It is worth noting that a similar single commodity flow-based model 
has been previously applied to ensure connected areas of biological significance in the context of Conservation Planning~\cite{Dilkina2010}.

More specifically, for a given graph $G = (V,E)$,
the \emph{C-STree} constraints use the augmented and oriented graph 
$G_0 = (V_0,A_0)$, where $V_0 = V \cup \{0\}$ and
$A_0 = A \cup \{(0,j): j \in V\}$, 
that is, $A_0$ includes the set $A$ and all
arcs from the new vertex $0$ (root) to all vertices in $V$.  
This characterization requires an additional set of variables $v_{ij}$ and $f_{ij}$, where $v_{ij}$ is set to one if the arc $(i,j)$ belongs to the spanning tree and zero otherwise. Similar to the MPR model, the variable $f_{ij}$ denotes the flow that passes through the arc $(i,j)$, for all $(i,j) \in A_0$.
Its characterization is given as follows.
 \begin{subequations}
 \begin{align}
    &\sum_{i:(i,j) \in A_0} v_{ij} = x_j & \forall \ j \in V\\
    &\sum_{j \in V} v_{0j}  =  1 \\
    & \sum_{i:(i,j) \in A_0} f_{ij} - \sum_{i:(j,i) \in A}f_{ji}  = x_j & \forall \ j \in V\\
    & f_{ij}  \geq  v_{ij} & \forall \ (i,j) \in A_0\\
    & f_{ij}  \leq  (u-1) v_{ij} & \forall \ (i,j) \in A\\
    & f_{0j}  \leq  u \cdot v_{0j} & \forall \ j \in V\\
    & \sum_{j \in V} f_{0j}  =  \sum_{j \in V}x_j\\
    &v_{ij} + v_{ji}  \leq  y_{ij} & \forall \ (i,j) \in A\\
    & v_{ij}  \in \{0,1\}    & \forall \ (i,j) \in A_0\\
    & f_{ij}  \geq 0    & \forall \ (i,j) \in A_0
    \end{align}
 \end{subequations}
Recall that the variables $x_j$ and $y_{ij}$ are the same defined in 
models F3 and M1 
for each vertex $j \in V$ and for each edge $\{i, j\} \in E$, respectively, and $u$ is an upper bound
on the size of the subgraph.
Constraints (5a) 
ensure that the spanning tree contains, for every selected node, exactly one arc that enters this node.
Constraint (5b)
ensures that the root vertex $0$ is
connected to exactly one vertex in the subgraph, 
from which all the flow is injected.
Constraints (5c)
model the flow balance on each vertex.
Constraints (5d)
prevent arc $(i,j)$ in the tree if it has no flow.
Constraints (5e) and (5f)  
force the arc in the tree
if there is flow passing through it. It also imposes an upper bound for that flow, being equal
to $u$ if it is the arc emanating from the root, and equal to 
$(u - 1)$ in all other cases.
In addition, Constraint (5g)
states that the flow sent from the root 
is exactly equal to
the vertex cardinality of the subgraph.
Constraints (5h)
impose that at most a single arc is chosen in the spanning tree if the edge $\{i,j\}$ is in the subgraph, and that the edge must be in the subgraph if there is an arc between the two vertices in the spanning tree. 
Finally, (5i)
characterize integrality conditions on the topological variables $v_{ij}$
and (5j) set non-negativity conditions on the flow variables $f_{ij}$.
Note that constraints (5d) and (5h) can be considered redundant,
but they strengthen the associated linear programming relaxation. 
C-STree requires the addition of 
$O(|E|)$ variables and 
$O(|V|+|E|)$
constraints to Model F3 and M1.

Inspired by the MPR constraints (3a-3i) proposed by Marinelli et al.~\cite{Marinelli2021}, we propose the \emph{C-Flow} constraints, which are designed to be included in Model M1 
to enforce connected quasi-cliques for DCKS using a single commodity flow-based approach. These constraints include a new binary variable $s_j$, where $s_j = 1$ if the vertex $j$
is selected as the source, and $0$ otherwise.
These variables represent the same as the $c_i$ and $v_{0j}$ variables used in
the MPR model and C-STree characterization, respectively. Similar to \emph{C-STree}, the variable $f_{ij}$ is used to denote the flow passing through the arc $(i,j)$ for all $(i,j)$ in $A$.
 \begin{subequations}
    \begin{align}
        &\sum_{j \in V} s_j = 1\\
    & s_j  \leq  x_j & \forall \ j \in V\\
    & f_{ij}  \leq  k \cdot y_{ij} & \forall \ \{i,j\} \in E\\
    & f_{ji}  \leq  k \cdot y_{ij} & \forall \ \{i,j\} \in E\\
    & \sum_{j:(i,j) \in E} f_{ji} - \sum_{j:(i,j) \in E}f_{ij}  =  x_i - k \cdot s_i & \forall \ i \in V\\
    & f_{ij}  \geq  0  & \forall \ (i,j) \in A\\    
    & s_{j} \in\{0,1\}    & \forall \ j \in V
    \end{align}
 \end{subequations}
Constraints (6a) and (6b)
ensure that only one vertex $j \in S$ is selected as the source. 
Constraints (6c) and (6d) set the bounds to the flow
variables $f_{ij}$, forcing edge $\{i,j\}$ in the solution if there is flow passing
between the two vertices. 
Constraints (6e)
model the flow balances at the vertices and enforce that exactly
$k-1$ units of flow leave the source vertex.
Finally, constraints (6g) 
enforce the source vertex variables to be binary. 
Due to the requirement of knowing the value of the parameter $k$, C-Flow can only be used in conjunction with model M1, where it adds $O(|E|)$ additional variables and 
$O(|V|+|E|)$ constraints.

\section{Computational Experiments}
\label{Res}

In order to assess the performance of C-STree and C-Flow in terms of run-time,
we used a set of $15$ real-life sparse graph instances obtained from the University of Florida Sparse Matrix Collection \cite{SparseMatrix}, along with the graph \emph{Homer} available at \href{https://mat.tepper.cmu.edu/COLOR/instances.html}{https://mat.tepper.cmu.edu/COLOR/instances.html}. 
The number of vertices, number 
of edges and edge density of these graphs are presented in Table \ref{tab:TestedGraphs}. 
All the graphs have been made undirected and simple by ignoring the direction of the arcs 
and removing self-loops and multiple edges. 
Additionally, when dealing with disconnected 
graphs  
(identified by the prefix \emph{Cp-} in Table~\ref{tab:TestedGraphs}), only the largest component
was taken into account.

For evaluation, C-STree was added to the F3 and M1 models, while the C-Flow was integrated into the M1 model. Both models enhanced with the proposed constraints were solved using the Gurobi Optimizer version 10.0.2 with the Python interface. 
The solver was configured with a thread count 
limit of $1$ (to not use multithreading), a time limit of $3\,600$ seconds, a memory constraint of 10GB, and the {\tt MIPGap} parameter set to $10^{-8}$.  

For comparison purposes, we implemented the MPR and Lazy constraints and also used the implementation of Connected$_\mu$ available at \href{http://fpt.akt.tu-berlin.de/connected-mu-clique}{http://fpt.akt.tu-berlin.de/connected-mu-clique} for the MCQC problem, and the FixCon approach available at \href{https://www.uni-marburg.de/en/fb12/research-groups/algorith/software/fixcon}{https://www.uni-marburg.de/en/fb12/research-groups/algorith/software/fixcon} for the DCKS.
The computational experiments were conducted on a computer cluster with two Intel Xeon Silver 4210R 2.4G
processors with 10 cores and 251GB of memory running under DebianGNU$\backslash$Linux 12 (bookworm).

\begin{table}[t] 
\centering
\caption{Characterization of the tested instances}
\label{tab:TestedGraphs}
\begin{tabular}{lrrr}
\toprule
              Graph & $|V|$  & $|E|$   & $dens$   \\ \hline
\midrule
         Cp-ca-GrQc &    4158&   13422 &  $<$0.01 \\
            Cp-geom &    3621&    9461 &  $<$0.01 \\
         Harvard500 &     500&    2043 &     0.02 \\
      Cp-netscience &     379&     914 &     0.01 \\
      Cp-California &    5925&   15770 &  $<$0.01 \\
           Polbooks &     105&     441 &     0.08 \\
           Cp-Homer &     542&    1619 &     0.01 \\
           Cp-yeast &    2224&    6609 &  $<$0.01 \\
             SmallW &     233&     994 &     0.04 \\
            USAir97 &     332&    2126 &     0.04 \\
        Cp-Erdos971 &     429&    1312 &     0.01 \\
 Celegans-metabolic &     453&    2025 &     0.02 \\
              Email &    1133&    5451 &     0.01 \\
             Cp-EVA &    4475&    4652 &  $<$0.01 \\
            Erdos02 &    5534&    8472 &  $<$0.01 \\
             As-735 &    6474&   12572 &  $<$0.01 \\
\bottomrule
\end{tabular}
\end{table}

In the following sections, we discuss the results 
obtained from our experiments.

\subsection{F3 and M1 models without connectedness constraints}

The goal of the first set of experiments is to measure the run-time of the MILP solver on the F3 and M1 models without
connectedness constraints and count how many 
disconnected quasi-cliques are found in the instances
described in Table~\ref{tab:TestedGraphs}. 
For the model F3, the parameter $\gamma$ was varied from $0.1$ to $1.0$, inclusively, with a step size of $0.01$, resulting in a total of $91$ tested values for each instance. 
For the model M1, the parameter $k$ was varied from 
two to the number 
of vertices minus one for each instance, inclusively, with a step size of one, resulting in a total of $|V|-2$ tested values for each instance.

\begin{table}[t]
\centering
\caption{Results for F3 and M1 models without connectedness constraints}
\label{tab:ResF3M1}
\begin{tabular}{lr@{  }r@{\quad}r@{ }r@{ }l@{\quad}r 
                 r@{  }r@{\quad}r@{ }r@{ }l}
\toprule
\multirow{2}{*}{Graph} & \multicolumn{5}{c}{F3 Model (MQC)} & & \multicolumn{5}{c}{M1 Model (DKS)}  \\
& $\%succ$ & \emph{$\%disc$} & \multicolumn{3}{c}{\emph{run-time}} & 
& $\%succ$ & \emph{$\%disc$} & \multicolumn{3}{c}{\emph{run-time}} \\ \hline
\midrule
         Cp-ca-GrQc &    100.0 &  74.7 & 444.9 & $\pm$& 535.7 &  &
                         100.0 &  30.1 &   7.4 & $\pm$&  50.9 \\
          Cp-geom   &    100.0 &   0.0 &  13.7 & $\pm$&  9.10 &&
                         100.0 &  21.9 &   1.3 & $\pm$&   5.6 \\
         Harvard500 &    100.0 &  76.9 &   8.2 & $\pm$&  11.5 &&
                         100.0 &  20.9 &   1.2 & $\pm$&   4.5 \\
      Cp-netscience &    100.0 &  38.5 &   0.1 & $\pm$&   0.1 &&
                         100.0 &  26.0 &   0.1 & $\pm$&   0.1 \\
      Cp-California &     44.0 &  10.0 & 744.3 & $\pm$&1074.8 &&
                          99.4 &   1.1 &  14.4 & $\pm$& 116.9 \\
           Polbooks &    100.0 &   2.2 &   0.3 & $\pm$&    0.2 &&
                         100.0 &   6.8 &   0.1 & $\pm$&    0.1 \\
              Cp-Homer & 100.0 &   3.3 &   4.1 & $\pm$&  4.6 &&
                         100.0 &   0.0 &   0.1 & $\pm$&  0.3  \\
              Cp-yeast & 100.0 &   2.2 &  29.1 & $\pm$&  52.8 &&
                         100.0 &   0.0 &   1.0 & $\pm$&  2.2 \\\hline
             SmallW    & 100.0 &   0.0 &   1.7 & $\pm$&  1.8 &&
                         100.0 &   0.0 &   0.1 & $\pm$&  0.4  \\
            USAir97 &     98.9 &   0.0 &  62.7 & $\pm$&  369.8  &&
                         100.0 &   0.0 &   0.4 & $\pm$&  1.6   \\
        Cp-Erdos971 &    100.0 &   0.0 &   6.5 & $\pm$&  7.0 &&        
                         100.0 &   0.0 &   0.2 & $\pm$&  0.5   \\
 Celegans-metabolic &    100.0 &   0.0 &   3.9 & $\pm$&  4.6 &&
                         100.0 &   0.0 &   0.2 & $\pm$&  0.5  \\
              Email &    100.0 &   0.0 & 540.9 & $\pm$&  320.1 &&
                         100.0 &   0.0 &  24.1 & $\pm$&  74.9    \\
             Cp-EVA &    100.0 &   0.0 &   1.5 & $\pm$&  1.1  &&
                         100.0 &   0.0 &   0.4 & $\pm$&  0.2   \\
            Erdos02 &    100.0 &   0.0 &  18.9 & $\pm$&  10.1  &&
                         100.0 &   0.0 &   0.5 & $\pm$&  0.8  \\
             As-735 &    100.0 &   0.0 &  18.5 & $\pm$&  22.9  &&
                         100.0 &   0.0 &   0.9 & $\pm$&  1.0   \\

\bottomrule
\end{tabular}
\end{table}

Table~\ref{tab:ResF3M1} presents the results obtained on all instances described in Table~\ref{tab:TestedGraphs} with both F3 and M1 models, considering all values of $\gamma$ for the F3 model and all values of $k$ for the M1 model. 
The first group of instances correspond to those for which at least
one optimal quasi-clique was found to be disconnected
in at least one of the models. In the second group, all
optimal quasi-cliques found were connected.
The column \emph{\%succ} shows the percentage of solved instances
within the time limit. The column $\%disc$ indicates the 
percentage of  
disconnected quasi-cliques obtained within the tested values
by each model for each instance. 
Column $run$-$time$
provides 
the average running-time and standard deviation, 
measured in seconds, required to solve 
each instance within the time limit averaged over all tested values for
each model.

The results indicate that the number of disconnected 
quasi-cliques can rise quite strongly, mainly
on \emph{Cp-ca-GrQc}, \emph{Harvard500}, and \emph{Cp-netscience}
instances for the F3 model and, in addition, on \emph{Cp-geom} 
for the M1 model. While comparing the running times between both models is not possible due to the disparity in parameter ranges and their implications for finding the solutions by each model,
it is noteworthy that the F3 model presents a high variance.

\subsection{Connected quasi-cliques for MCQC}

This section reports the results for the MCQC Problem obtained from the 
F3 model using the \emph{MPR} constraints (F3+MPR) 
and the \emph{C-STree} constraints (F3+C-STree), 
as well as the Connected$_{\mu}$ approach.
We have considered only the instances in Table~\ref{tab:TestedGraphs}
that have produced disconnected quasi-cliques
in the previous experiment for the F3 model
(see instances with positive values in column $\%disc$ in Table~\ref{tab:ResF3M1}).
The experiments were conducted for different $\gamma$ values, specifically, $\gamma \in \{0.1, 0.2, \ldots, 0.9\}$.

Table~\ref{tab:ResMQC} presents the results for the three approaches over all values of $\gamma$ as previously defined.
Column $\%succ$ corresponds to the percentage of instances solved within
the time limit, and column \emph{run-time} corresponds to the average
and standard deviation of running time in seconds for each approach 
and each instance.
The results are divided into two groups: the first group corresponds to cases with $\gamma < 0.5$, and the second group is related to cases with $\gamma \geq 0.5$. It is important to note that we only present results for the Connected$_\mu$ approach in the second group (i.e., for $\gamma \geq 0.5$) because this method cannot effectively solve the problem for $\gamma < 0.5$. 
In the table, MLE means that 
an approach was never able to solve that instance for any value of 
$\gamma$ within the memory limit.

For the first group of results ($\gamma < 0.5$), our approach (F3+C-STree) demonstrates superior performance, achieving $100\%$ success rate on all graphs, except for \emph{Cp-ca-GrQc}, where it achieves a still noteworthy $50\%$. Additionally, it outperforms the F3+MPR model in terms of run-time across the majority of the graphs. 
It should be highlighted that when $\gamma < 0.5$, instances tend to exhibit a higher number of disconnected quasi-cliques. This was observed in $66.8\%$ of cases for which the F3 model reported disconnected solutions. These instances precisely align with the cases wherein F3+C-STree demonstrates better performance.

For the second group of results,  our approach is shown to be very competitive. 
The Connected$_{\mu}$ algorithm shows limited success and can only solve very few instances. 
Although it is the fastest method for the graphs \emph{Harvard500}, \emph{Cp-yeast}, and \emph{Cp-California},
it manages to solve only $40\%$ of the instances for these graphs. 
In contrast, both MILP models were successful in solving $100\%$ of the instances for these graphs, except for \emph{Cp-California}, which also presented challenges.
Additionally, the graph \emph{Cp-ca-GrQc} was proven to be challenging for all approaches,
with F3+C-STree achieving the best performance, being able to solve 
$60\%$ of the instances. 
On the remaining graphs, all the approaches successfully solved all the instances, but the MILP models demonstrated the fastest performance, on average.
In general, both the MILP models for the connected cases are slower than the F3 model (see Table~\ref{tab:ResF3M1}), which is due to the additional constraints added to the model.

Additional experiments were conducted to compare the size of quasi-cliques produced by approaches that ensure connectedness versus those that do not. To achieve this, we ran the connectedness-guaranteed approaches with the same $\gamma$ parameter values that resulted in disconnected solutions in the F3 model. The results show that the size of quasi-cliques produced by both approaches differs for specific graph instances and gamma values. In particular, this was observed in graphs Cp-ca-GrQc, Cp-netscience, and Harvard500, which presented 2, 15, and 28 cases, respectively. In the worst-case scenario, observed in experiments with the 
graph
Harvard500,
the size of the maximum connected quasi-clique was up to $5\%$ smaller than when not considering connectedness. 

\begin{table}[t]
\centering
\caption{Results for MCQC Problem using the Connected$_\mu$ approach,
\emph{F3} Model with MPR constraints (F3+MPR) and \emph{F3} Model with 
\emph{C-STree} constraints (F3+C-STree).}
\label{tab:ResMQC}
\setlength\tabcolsep{1.5pt} 
\begin{tabular}{r@{\quad} r@{ }
r@{ }r@{ }r@{ }l@{ }c@{\quad}
r@{ }r@{ }r@{ }l@{ }c@{\quad} 
r@{ }r@{ }r@{ }l@{ }}
\toprule
        \multicolumn{1}{c}{\multirow{2}{*}{Graph}} &  
        \multicolumn{1}{c}{\multirow{2}{*}{$\gamma$}} &       
        \multicolumn{4}{c}{Connected$_\mu$} & &  
        \multicolumn{4}{c}{F3+MPR} & &   
        \multicolumn{4}{c}{F3+C-STree} \\
        \multicolumn{1}{c}{} &
        \multicolumn{1}{c}{} & 
        \multicolumn{1}{c}{$\%succ$} &
        \multicolumn{3}{c}{\emph{run-time}} & &
        \multicolumn{1}{c}{$\%succ$} &
        \multicolumn{3}{c}{\emph{run-time}} & &
        \multicolumn{1}{c}{$\%succ$} &
        \multicolumn{3}{c}{\emph{run-time}} \\
\midrule
Polbooks
& \multicolumn{1}{c}{\multirow{7}{*}{$< 0.5$}} 
&     - & \multicolumn{3}{c}{-}  &
&  \bf{100.0} & \bf{0.4} & $\pm$ & 0.4 &
& \bf{100.0} &   0.9 & $\pm$ & 0.7   \\
Cp-netscience
& \multicolumn{1}{c}{}
&     - & \multicolumn{3}{c}{-}  &
& \bf{100.0} & \bf{11.0} & $\pm$ & 17.7 &
& \bf{100.0} &   15.1 & $\pm$ & 11.6   \\
Harvard500
& \multicolumn{1}{c}{}
&     - & \multicolumn{3}{c}{-}  &
&  \bf{100.0} & 40.7 & $\pm$ & 74.7  &
& \bf{100.0} &   \bf{28.7} & $\pm$ & 32.8   \\
Cp-Homer
& \multicolumn{1}{c}{}
&     - & \multicolumn{3}{c}{-}  &
& \bf{100.0} & 3.7 & $\pm$ & 2.0 &
& \bf{100.0} &   \bf{3.0} & $\pm$ & 0.5   \\
Cp-yeast
& \multicolumn{1}{c}{}
&     - & \multicolumn{3}{c}{-}  &
&   \bf{100.0} &   286.8 & $\pm$ & 87.5  &
& \bf{100.0} &   \bf{75.0} & $\pm$ & 27.8   \\ 
Cp-California 
&  \multicolumn{1}{c}{}
&     - & \multicolumn{3}{c}{-}  &
&   0.0 & \multicolumn{3}{c}{MLE}  &
& \bf{100.0} &   \bf{912.4} & $\pm$ & 1\,076.9   \\

Cp-ca-GrQc
& \multicolumn{1}{c}{}
&     - & \multicolumn{3}{c}{-}  &
&   0.0 & \multicolumn{3}{c}{MLE}  &
& \bf{50.0} &   \bf{2\,594.8} & $\pm$ & 166.3   \\ \hline

Polbooks
& \multicolumn{1}{c}{\multirow{7}{*}{$\geq0.5$}}   
&  \bf{100.0} &   3.4 & $\pm$ & 5.6 &
&  \bf{100.0} &   \bf{1.1} & $\pm$ & 1.3  &
&  \bf{100.0} &   3.3 & $\pm$ & 4.5   \\  
Cp-netscience
& \multicolumn{1}{c}{}
&  \bf{100.0} &   47.2 & $\pm$ & 91.3  &
&  \bf{100.0} &   \bf{0.8} & $\pm$ & 0.2 &
&  \bf{100.0} &   \bf{0.8} & $\pm$ & 0.2   \\
Harvard500
& \multicolumn{1}{c}{}
&  40.0 &   \bf{51.1} & $\pm$ & 50.1   &
&  \bf{100.0} &   276.5 & $\pm$ & 710.6 & 
&  \bf{100.0} &   129.4 & $\pm$ & 115.0  \\
Cp-Homer
& \multicolumn{1}{c}{}
&   40.0 &   12.4 & $\pm$ & 12.1  &
&  \bf{100.0} &   \bf{9.9} & $\pm$ & 9.3 &
& \bf{100.0} &   46.0 & $\pm$ & 55.9   \\
Cp-yeast
& \multicolumn{1}{c}{}
&  40.0 &   \bf{9.6} & $\pm$ & 8.5  &
&  \bf{100.0} &   354.3 & $\pm$ & 48.2 &
&  \bf{100.0} &   201.9 & $\pm$ & 78.3   \\
Cp-California 
& \multicolumn{1}{c}{}
&  \bf{40.0} & \bf{182.9} & $\pm$ & 175.5 &
&  0.0 & \multicolumn{3}{c}{MLE} &
& 20.0 &   3\,486.1 & $\pm$ & 0.0   \\
Cp-ca-GrQc
& \multicolumn{1}{c}{}
&  20.0 &   1\,784.8 & $\pm$ & 0.0  &
&   0.0 & \multicolumn{3}{c}{MLE}  &
& \bf{60.0} &   \bf{696.2} & $\pm$ & 315.5   \\
\bottomrule
\end{tabular}
\end{table}

\subsection{Connected quasi-cliques for DCKS}

This section presents the outcomes for DCKS Problem obtained using the M1 Model with the \emph{Lazy} (M1+Lazy), \emph{C-STree} (M1+C-STree) and \emph{C-Flow} (M1+C-Flow) constraints, along with the FixCon approach. The results are summarized in Table~\ref{tab:ResDCKS}, which follows a similar format as Table~\ref{tab:ResMQC}. These outcomes correspond to varying the parameter $k$ from two to the number of vertices minus one for each instance.
The results are organized into two distinct categories: the first one covers cases where $k \leq 20$, and the second category relates to cases where $k > 20$. Due to its limitations in handling instances with $k > 20$, the outcomes of the FixCon approach are exclusively presented within the first category, corresponding to cases where $k\leq20$.

For instances where $k\leq20$, the M1+Lazy model achieves a $100\%$ success rate for most graphs with competitive run-times. 
These findings are not surprising because the chances of having disconnected subgraphs are lower when the value of $k$ is small. As a result, the need to execute the lazy constraints to ensure connectedness occurs less frequently.
Although FixCon presents better run-times for most graphs, it shows lower success rates compared to the other three approaches, indicating a comparatively weaker performance. In contrast, our proposed methods, M1+C-STree and M1+C-Flow, show competitive outcomes both in terms of success rate and run-times. 

For the second category
with $k>20$, M1+C-Flow shows better performance, achieving a $100\%$ success rate for the majority of instances within a quite competitive run-time, although it might have slightly higher run-times compared to M1+Lazy in some graphs such as Polbooks, Harvard500, and Cp-ca-GrQc. Both Cp-California and Cp-ca-GrQc graphs impose challenges for the three approaches with M1+C-Flow being able to solve more instances. 
It is worth observing that, in our experimental context composed of real-life sparse graphs, the majority of disconnected solutions emerge when $k > 20$. The M1 model results confirm that $99.9\%$ of the disconnected solutions relate to instances where $k$ exceeds $20$. It is precisely within this subset of cases that our proposed constraints demonstrate 
better performance. 
 
We also have compared the density of quasi-cliques achieved by approaches ensuring connectedness against those that do not. For this comparison, we ran the connectedness-guaranteed approaches with the same $k$ parameter values that resulted in disconnected solutions in the M1 model. The results from these experiments show that for all the tested instances, except Polbooks, the density of quasi-cliques differs between the two approaches. This was especially noticeable in graphs Cp-ca-GrQc, Cp-netscience, Harvard500, Cp-geom, and Cp-California, which presented 625, 93, 101, 757, and 21 cases, respectively. In the worst-case scenario, observed in experiments with Cp-netscience, the density of the maximum connected quasi-clique was up to $4,6\%$ smaller than when not considering connectedness.

\begin{table}[t]
\footnotesize
\centering
\caption{Results for the DCKS Problem using the FixCon approach,
\emph{M1} Model with \emph{Lazy} (M1+Lazy), \emph{C-STree} (M1+C-STree), and \emph{C-Flow} (M1+C-Flow) constraints.}
\label{tab:ResDCKS}
\setlength\tabcolsep{0.2pt} 
\begin{tabular}{r@{\hspace{0.2em}} r@{ }
r@{ }r@{ }r@{ }l@{ }c@{\hspace{0.2em}}
r@{ }r@{ }r@{ }l@{ }c@{\hspace{0.2em}}
r@{ }r@{ }r@{ }l@{ }c@{\hspace{0.2em}} 
r@{ }r@{ }r@{ }l@{ }}
\toprule
        \multicolumn{1}{c}{\multirow{2}{*}{Graph}} &  
        \multicolumn{1}{c}{\multirow{2}{*}{$k$}} &       
        \multicolumn{4}{c}{FixCon} & & 
        \multicolumn{4}{c}{M1+Lazy} & &   
        \multicolumn{4}{c}{M1+C-STree} & &   
        \multicolumn{4}{c}{M1+C-Flow} \\
        \multicolumn{1}{c}{} &
        \multicolumn{1}{c}{} & 
        \multicolumn{1}{c}{$\%succ$} &
        \multicolumn{3}{c}{\emph{run-time}} & &
        \multicolumn{1}{c}{$\%succ$} &
        \multicolumn{3}{c}{\emph{run-time}} & &
        \multicolumn{1}{c}{$\%succ$} &
        \multicolumn{3}{c}{\emph{run-time}} & &
        \multicolumn{1}{c}{$\%succ$} &
        \multicolumn{3}{c}{\emph{run-time}} \\
\midrule
Polbooks
& \multicolumn{1}{c}{\multirow{7}{*}{$\leq 20$}} 
&   73.7 & 406.1 & $\pm$ & 977.1 &
&  \bf{100.0} & \bf{0.2} & $\pm$ & 0.1  &
&  \bf{100.0} & 1.1 & $\pm$ & 1.1 &
&  \bf{100.0} & 0.5   & $\pm$ & 0.3   \\
Cp-netscience
& \multicolumn{1}{c}{}
&   78.9 & 230.0 & $\pm$ & 678.5 &
& \bf{100.0}  & \bf{0.4} & $\pm$ & 0.5 &
& \bf{100.0} & 1.4 & $\pm$ & 1.8 &
& \bf{100.0} & 0.8   & $\pm$ & 0.9   \\
Harvard500
& \multicolumn{1}{c}{}
&  \bf{100.0} & \bf{0.5} & $\pm$ & 0.2 &
&  \bf{100.0} & 0.9 & $\pm$ & 0.6 &
&  \bf{100.0} & 5.6 & $\pm$ & 3.5  &
&  \bf{100.0} & 2.5  & $\pm$ & 1.1   \\
Cp-geom
& \multicolumn{1}{c}{}
& 84.2 & \bf{2.5} & $\pm$ & 5.6 &
& \bf{100.0}  & 96.8 & $\pm$ & 195.9 &
& 89.5 & 170.5 & $\pm$ & 237.4 &
& 94.7 & 145.4  & $\pm$ & 363.2   \\

Cp-California 
&  \multicolumn{1}{c}{}
&   57.9 & \bf{175.0} & $\pm$ & 266.4 &
&  \bf{73.7}  & 811.8 & $\pm$ & 921.7 &
&   47.4 & 1\,661.8 & $\pm$ & 565.9 &
&  63.2 & 761.8   & $\pm$ & 302.6   \\

Cp-ca-GrQc
& \multicolumn{1}{c}{}
&  \bf{100.0} & \bf{0.7} & $\pm$ & 0.1 &
&  \bf{100.0}  & 231.2 & $\pm$ & 421.7 &
&   84.2 & 376.3 & $\pm$ & 565.7 &
& 89.5 & 376.0  & $\pm$ & 655.6  \\ \hline

Polbooks
& \multicolumn{1}{c}{\multirow{7}{*}{$> 20$}}   
&     - & \multicolumn{3}{c}{-}  &
&  \bf{100.0} & \bf{0.1} & $\pm$ & 0.1 &
&  \bf{100.0} & 0.3 & $\pm$ & 0.4  &
&  \bf{100.0} & 0.2   & $\pm$ & 0.2  \\  
Cp-netscience
& \multicolumn{1}{c}{}
&     - & \multicolumn{3}{c}{-}  &
&   \bf{100.0} & 18.7 & $\pm$ & 82.7 &
&  \bf{100.0} &   15 & $\pm$ & 56.8 &
&  \bf{100.0} & \bf{3.2}   & $\pm$ & 8.5   \\
Harvard500
& \multicolumn{1}{c}{}
&     - & \multicolumn{3}{c}{-}  &
&  \bf{100.0} & \bf{2.8} & $\pm$ & 12.8 &
&  \bf{100.0} &   11.9 & $\pm$ & 37.7 & 
&  \bf{100.0} & 3.0   & $\pm$ & 10.8  \\
Cp-geom
& \multicolumn{1}{c}{}
&     - & \multicolumn{3}{c}{-}  &
&  90.2  & 85.3 & $\pm$ & 317.9 &
&  99.9 &   218.0 & $\pm$ & 322.0 &
&  \bf{100.0} & \bf{46.2}   & $\pm$ & 178.6   \\
Cp-California 
& \multicolumn{1}{c}{}
&     - & \multicolumn{3}{c}{-}  &
&  98.4  & 36.0 & $\pm$ & 164.7 &
&  98.4 &   181.9 & $\pm$ & 256.9 &
&  \bf{99.2} & \bf{31.3}   & $\pm$ & 163.5    \\
Cp-ca-GrQc
& \multicolumn{1}{c}{}
&     - & \multicolumn{3}{c}{-}  &
&  91.5  & \bf{88.0} & $\pm$ & 261.2 &
&   93.3 &   318.3 & $\pm$ & 398.1 &
& \bf{93.7} & 112.2   & $\pm$ & 446.1   \\
\bottomrule
\end{tabular}
\end{table}

\section{Conclusions}
\label{Con}

In this paper, we have presented two different sets of constraints designed to ensure the connectedness of quasi-cliques. The first, C-STree, uses
a spanning tree representation based on the single commodity flow model and applies to both the MCQC and DCKS problems when integrated into the F3 and M1 models, respectively. The second, C-Flow, uses a classic flow-based approach and is specifically applicable to the DCKS problem when integrated into the M1 model.

Through extensive empirical analysis, we have provided a comprehensive evaluation of these proposed constraints, comparing them with the Connected$_{\mu}$ algorithm and the MPR model for the MCQC Problem, and 
with the FixCon approach and the Lazy model for the DCKS Problem. 
The results show that our proposed C-STree constraints integrated with the F3 model emerged as the most effective choice, in general, among the evaluated methods to solve the MCQC, especially for cases where the density parameter $\gamma$ is below $0.5$. 
In the context of the DCKS Problem our proposed C-Flow constraints integrated with the M1 model proved to be very competitive for solving instances where $k\leq20$ while it emerges as the best choice for cases where $k>20$. In conclusion, C-STree and C-Flow show the best performance precisely when disconnected quasi-cliques become more prevalent, i.e. for larger values of $k$ and smaller values of $\gamma$.  
It is noteworthy to emphasize a significant characteristic shared by both of our proposed methods, C-STree and C-Flow: their inherent lack of limitations on parameter values. This versatility enhances their applicability in a broad range of scenarios. 

In a future work, we intend to contribute to the field by exploring strategies that ensure the connectedness property on heuristic approaches.

\backmatter

\bmhead{Acknowledgments}
D.~S dos Santos acknowledges the Foundation for Science and Technology (FCT)
for the Ph.D.~fellowship 2022.12082.BD. This work is partially
funded by the FCT, 
I.P./MCTES through 
national funds (PIDDAC), 
within the scope of CISUC R\&D Unit - UID/CEC/00326/2020.

\section*{Declarations}
\subsection{Availability of data and materials}
This work has no associated data.
\subsection{Conflict of interest}
All authors have no conflicts of interest.

\hide{
\subsection{Funding}
\subsection{Consent to participate}
Not applicable
\subsection{Consent for publication}
Not applicable
\subsection{Code availability}
Not applicable
\subsection{Authors' contributions}
Not applicable
}

\bibliography{References}

\end{document}